\documentstyle[preprint,aps,epsfig]{revtex}
\tightenlines
\begin{document}
\draft
%%%%%%%%%%%%%%%%%%%%%%%%%%%%%%%%%%%%%%%%%%%%%%%%%%%
%%%%%%%%%%%% Begin Cover Page %%%%%%%%%%%%%%%%%%%%%%%%%%%%%%%%%%%%%%%%%%
\preprint{\begin{minipage}[b]{1.5in}
          CU-TP-925\\
          UK/TP 98-18
          \end{minipage}}
\title{Gluon mini-jet production in nuclear collisions at high energies}
\author{Xiaofeng Guo}
\address{ Department of Physics and Astronomy,
         University of Kentucky,\\
         Lexington, KY 40506, U. S. A.\\
          and \\
         Department of Physics, 
         Columbia University,\\
         New York, NY 10027, U. S. A.  }
\date{November 28, 1998}
\maketitle

\begin{abstract} 
We show explicitly that the leading soft gluon $p_T$ distribution, 
predicted by Kovner, McLerran, and Weigert after solving classical 
Yang-Mills equations, can be understood in terms of conventional 
QCD perturbation theory. We also demonstrate that the key logarithm  
in their result represents the logarithm in DGLAP evolution equations.
\end{abstract} 
\vspace{0.2in}

%%%%%%%%%%%% End of Cover Page %%%%%%%%%%%%%%%%%%%%%%%%%%%%%%%%%%%%%%%%%

%%%%%%%%%%%%%% Begin Section I %%%%%%%%%%%%%%%%%%%%%%%%%%%%%%%%%%%%%%%%%
\section{Introduction}

In ultra-relativistic heavy ion collisions, physical observables
sensitive to a few GeV momentum scale, such as the mini-jet
production, will be dominated by scattering of soft gluons from both
heavy ion beams. Understanding the distribution of soft
gluons formed in the initial stage of the collision is particularly
interesting and important.  In terms of conventional QCD perturbation
theory, a calculable cross
section in high energy hadronic collisions is factorized into a single 
collision between two partons multiplied by a probability to find
these two partons of momentum fractions $x_1$ and $x_2$, respectively,
from two incoming hadrons.  The probability is then factorized into a
product of two parton distributions $\phi(x_1)$ and $\phi(x_2)$, 
which are probabilities to find these two partons from the respective 
hadrons \cite{Factorization}.
Because of extremely large number of soft gluons in heavy 
ion beams, it is natural to go beyond
the factorized single-scattering formalism to include any possible
multiple scattering, and long range correlations between soft gluons
from two incoming ions.

Recently, McLerran and Venugopalan (MV) developed a new formalism for 
calculation of the soft gluon distribution for very large 
nuclei \cite{Raju1,Raju2}. In this approach, the valence quarks in the 
nuclei are treated as the classical source of the color charges.
They argued that the valence quark recoil can be ignored in the limit
when the gluons emitted are soft.  
In addition, because of the Lorentz contraction, the color charge 
of the valence quarks is treated approximately as an infinitely 
thin sheet of color charge along the light cone. 
With these assumptions, the gluon distribution function for very large 
nuclei may be obtained by solving the classical Yang-Mills Equation
\cite{Raju2,Raju3}. Using the classical glue field generated by a
single nucleus obtained in the MV formalism as the basic
input, Kovner, McLerran, and Weigert (KMW)
computed the soft gluon production in a collision of two 
ultra-relativistic heavy nuclei by solving the classical  
Yang-Mills equations with the iteration to the second order \cite{KMH}. 
The two nuclei are treated as the infinitely thin sheets of the 
classical color charges  moving at the speed of light in the positive 
and the negative $z$ directions, respectively.   Following this 
approach, the distribution of soft gluons at the rapidity $y$ and the
transverse momentum $p_T$ in nuclear collisions can be express as
\cite{KMH}
\begin{equation}
\frac{dN}{dyd^2 p_T}=S_T\, \frac{2g^6 \mu^4}{(2\pi)^4}\, 
 N_c(N_c^2-1)\, \frac{1}{p_T^4}\, 
 \ell n \left(\frac{p_T^2}{\Lambda^2_{cutoff}}\right) \ ,
\label{e1} 
\end{equation}
where $g$ is the strong coupling constant, $N_c=3$ is the number
of the color, and $\Lambda_{cutoff}$ is a cutoff mass scale \cite{KMH}.  
Note that there was a factor of $\pi$ misprint in Eq.~(50) of 
Ref.~\cite{KMH}, as pointed out in Ref.~\cite{MGLM}.  
In Eq.~(\ref{e1}), $\mu^2$ is the
averaged color charge squared per unit area of the valence quark, and
$S_T$ is the transverse area of the nuclei.  
The $\mu^2$ and the $S_T$ are related as \cite{MGLM} 
\begin{equation}
S_T \, \mu^2=\frac{N_q}{2N_c} \ ,
\label{stmu}
\end{equation}
where $N_q$ is the number of valence quarks in the color charge
source.  The number distribution in Eq.~(\ref{e1}) can be also
expressed in terms of the cross section \cite{MGLM}
\begin{equation}
\frac{d \sigma}{dy d^2 p_T}=
             \frac{2g^6}{(2\pi)^4}
             \left(\frac{N_q}{2N_c}\right)^2 
              N_c (N_c^2-1)\, \frac{1}{p_T^4}\, 
    \ell n \left(\frac{p_T^2}{\Lambda^2_{cutoff}}\right)\ .
\label{ae0}
\end{equation}
In deriving Eq.~(\ref{ae0}), the following relation was used \cite{MGLM}
\begin{equation}
\frac{d \sigma}{dy d^2 p_T}=S_T\, \frac{dN}{dyd^2 p_T} \ .
\label{ae00}
\end{equation} 
The key result derived in Ref.~\cite{KMH}, Eq.~(\ref{e1}) (or 
equivalently, Eq.~(\ref{ae0})), is potentially very useful in 
estimating the production of mini-jet rates, and the formation 
of the possible quark-gluon plasma at RHIC \cite{GW}.
The purpose of this paper is to understand the respective role
of perturbative and non-perturbative QCD in deriving the expression
in Eq.~(\ref{e1}) (or that in Eq.~(\ref{ae0})), and explore under
what kind of approximation this result matches the conventional 
perturbative calculation.  

\begin{figure}
\epsfig{figure=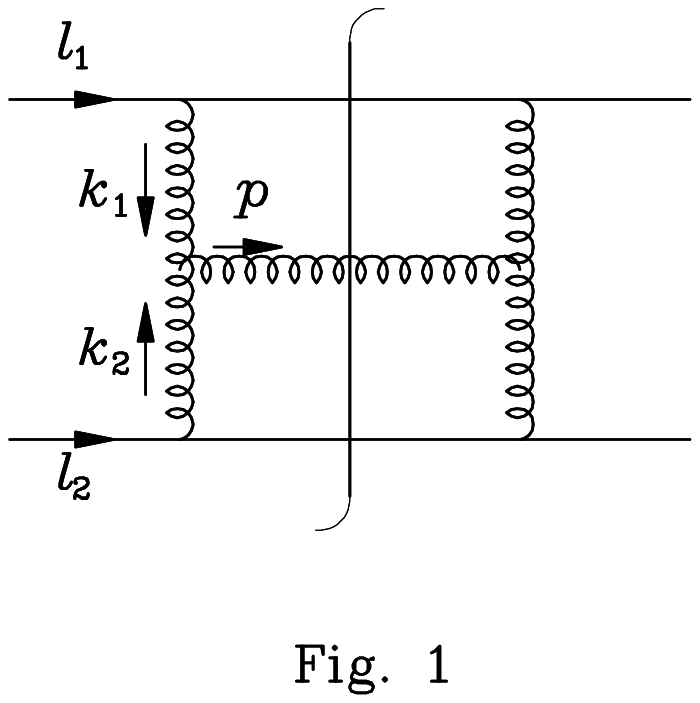,width=2.0in}
\caption{Square of the leading Feynman diagram to the process: 
$qq\rightarrow qqg$.}
\label{fig1}
\end{figure}

KMW's derivation for Eq.~(\ref{e1}) is based on the following physical 
picture: in ultra-relativistic heavy ion collisions, gluons are 
produced by the fields of two strongly Lorentz contracted color 
charge sources, which are effectively equal to the valence quarks of
two incoming ions.  In order to understand KMW's result in terms of
the language of perturbative QCD, we consider a specific partonic 
subprocess: $qq\rightarrow qqg$, as sketched in Fig.~\ref{fig1}.
If we assume that the incoming quarks $qq$ are the valence quarks in
the initial color charge sources, the partonic subprocess in
Fig.~\ref{fig1} mimics the physical picture adopted in KMW's
derivation.  However, as a Feynman diagram in QCD perturbation theory,
the single diagram shown in Fig.~\ref{fig1} is not gauge invariant.
As we demonstrate in Sec.~\ref{sec:2}, under certain
approximations, the contributions {\it extracted} from the diagram in
Fig.~\ref{fig1} to the leading soft gluon production in Eq.~(\ref{e1})
is gauge invariant; and therefore, the physical picture proposed by
KMW for soft gluon production is preserved.  

In Sec.~\ref{sec:3}, within the framework of 
conventional perturbative QCD, we
calculate the gluon production through the partonic subprocess
$qq\rightarrow qqg$, as shown in Fig.~\ref{fig1}, at the soft gluon
limit.  With our explicit calculation of this subprocess, we
demonstrate that Eq.~(\ref{e1}) (or the cross section in
Eq.~(\ref{ae0})) at $N_q=1$ can be reproduced.
Through our derivation, we show that the key logarithm 
$\ell n (p_T^2/\Lambda^2_{cutoff})$ in Eq.~(\ref{e1}) (or in 
Eq.~(\ref{ae0})) is basically the logarithm from the 
splitting of the incoming quark to the soft gluon in Fig.~\ref{fig1}.
In terms of the conventional QCD factorization formalism
\cite{Factorization}, such logarithm is normally factorized into the
distributions of the collinear gluons inside the incoming hadrons, and
the logarithmic dependence of the distributions is a direct result of
solving the DGLAP evolution equations \cite{GLAP}.  

Finally, in Sec.~\ref{sec:4}, we discuss the relations between the MV
formalism and that of the conventional QCD factorization.  We 
explicitly demonstrate that KMW's result can be reduced to the 
factorized formula in the conventional perturbative QCD, if we  
replace the charge density for the classical color charge 
$\mu^2$ (or equivalently $N_q$) by the valence quark distributions
of the nuclei, and absorb the logarithm 
$\ell n (p_T^2/\Lambda^2_{cutoff})$ into 
one of the valence quark distributions.  We point out that 
with the higher order of iteration, KMW's approach may include the 
multi-parton dynamics which is not apparent in the conventional 
perturbative calculation.

%%%%%%%%%%%%%% End of Section I %%%%%%%%%%%%%%%%%%%%%%%%%%%%%%%%%%%%%%%%

%%%%%%%%%%%%%% Begin Section II %%%%%%%%%%%%%%%%%%%%%%%%%%%%%%%%%%%%%%%%
\section{Factorization and Gauge Invariance}
\label{sec:2}

As we discussed above, the partonic process $qq\rightarrow qqg$, 
as shown in Fig.~\ref{fig1}, mimics the physical picture adopted in 
KMW's derivation, if we assume that the incoming quarks $qq$ are the
valence quarks in the initial color charge sources.
However, in general, the Feynman diagram shown in 
Fig.~\ref{fig1} is not gauge invariant by itself. In this section, 
we discuss how to extract the gauge invariant leading contribution 
from the diagram in Fig. 1, and what kind of approximation we need 
to take in order to extract such leading contributions. 

For the production of gluons, we evaluate the invariant cross section,
$d\sigma_{qq\rightarrow g} / dy d^2 p_T$, with $y$ and $p_T$ the
repidity and the transverse momentum of the produced gluon.
We label $l_1$ and $l_2$ as the momenta of the two incoming quarks,
respectively, and we choose $k_1$ and $k_2$ to be the momenta for the two  
gluons emitted from the initial quarks.  We have $p^2=0$ for the
final-state gluon because of its on-shell condition.  For the other two
gluons, $k_1$ and $k_2$ can not be on shell at the same time, because
$k_1$ and $k_2$ come from different directions, and $p^2=(k_1+k_2)^2=0$.

In general, the cross section can be written as
\begin{equation}  
d\sigma = \frac{1}{2s}\, |\overline{M}|^2\, dps \ ,
\label{e2}
\end{equation}
where $s=(l_1+l_2)^2$, and $|\overline{M}|^2$ is matrix
element square with the initial-spin averaged and the final-spin summed.
In Eq.~(\ref{e2}), $dps$ is the phase space, and can be expressed as:
\begin{eqnarray}
dps &=& \frac{d^4 k_1}{(2\pi)^4} (2\pi) \delta ((l_1-k_1)^2-m^2)
    \times \frac{d^4 k_2}{(2\pi)^4} (2\pi) \delta ((l_2-k_2)^2-m^2)
 \nonumber \\
& &  \times \frac{d^3 p}{(2\pi)^3 2E} (2\pi)^4 \delta^4 (k_1+k_2-p)
\nonumber \\
&=& \frac{d^3 p}{E} \frac{1}{2(2\pi)^3} \frac{d^4 k_1}{(2\pi)^4}
   (2\pi)^2 \delta ((l_1-k_1)^2-m^2) \delta ((l_2-k_2)^2-m^2) \ ,
\label{e3}
\end{eqnarray}
where $k_2=p-k_1$, and $m$ is the quark mass. For simplicity, we assume 
that both incoming quarks have the same mass. In high energy collisions, 
we set the mass of light quarks to be zero. Because of the 
gluon propagators, as shown in Fig.~\ref{fig1}, the matrix element
square $|\overline{M}|^2$ has the following pole structure:  
\begin{equation}
poles=\frac{1}{k_1^2+i\epsilon}\, \frac{1}{k_1^2-i\epsilon}\,
      \frac{1}{k_2^2+i\epsilon}\, \frac{1}{k_2^2-i\epsilon}\, .
\label{e4}
\end{equation}
When integrating over the phase space, we see that the leading
contribution comes from the terms with $k_1^2 \rightarrow 0$ or
$(p-k_1)^2=k_2^2 \rightarrow 0$ limit. 
As pointed out above, $k_1^2$  and $k_2^2$ can not be zero at the
same time.  Therefore, to calculate the leading contribution, we can
first calculate the diagram in $k_1^2 \rightarrow 0$ limit.  The total
leading contribution is just twice of it, because the diagram is
symmetric for $k_1$ and $k_2$. 

When we take $k_1^2 \rightarrow 0$, the integration become divergent. 
Therefore, an introduction of a cutoff is necessary for obtaining a 
finite contribution from the diagram in Fig.~\ref{fig1}, and the 
corresponding contributions are sensitive to the cutoff.  To derive 
the leading contribution at $k_1^2 \rightarrow 0$ limit, 
we perform the collinear approximation
$k_1\approx xl_1+O(k_{1T})$, with $k_{1T} \sim \Lambda_{cutoff} <<
p_T$, where $\Lambda_{cutoff}$ is a collinear cutoff scale 
\cite{GS:TASI}.  This approximation means that the leading 
contribution is from the phase space where almost all 
transverse momentum of the final-state gluon comes from the gluon of
$k_2$, and $k_1$ is almost collinear to $l_1$.  After such collinear 
approximation, the cross section in Eq.~(\ref{e2}) can be 
approximately written in a factorized form \cite{Factorization}: 
\begin{equation}
E \frac{d\sigma_{qq\rightarrow g}}{d^3 p} 
    \approx 2\left(\frac{1}{2(2\pi)^3}\, \frac{1}{2s} \right)
            \int \frac{dx}{x} \, 
             P_{l_1 \rightarrow k_1} (x, k_{1T}<p_T) \, 
             H(xl_1,l_2,p)
         +  O(\frac{\Lambda_{cutoff}^2}{p_T^2})  \, ,
\label{e5}
\end{equation}
where the overall factor of 2 is due to the fact that  the leading 
contribution come from two regions corresponding to 
$k_1^2\rightarrow 0$ and $k_2^2\rightarrow 0$, respectively. 
In Eq.~(\ref{e5}), $P_{l_1 \rightarrow k_1}(x, k_{1T}<p_T)$ represents
the probability of finding an almost collinear gluon with the momentum
fraction $x$ from an incoming quark of momentum $l_1$, and 
\begin{equation}
P_{l_1 \rightarrow k_1} (x, k_{1T}<p_T) = 
       \int \frac{d^4 k_1}{(2\pi)^4} \,
            x\, \delta (x-\frac{k_1}{l_1})\, 
             |\overline{M}_{q \rightarrow g}|^2 \,
             (2\pi) \delta ((l_1-k_1)^2-m^2) \, .
\label{e6}
\end{equation}
The diagram for $|\overline{M}_{q \rightarrow g}|^2$ can be
represented by Fig.~\ref{fig2}.  $H(xl_1,l_2,p)$ in Eq.~(\ref{e5})
is effectively the hard scattering between the gluon of $k_1=xl_1$ 
and the incoming quark of $l_2$. and given by
\begin{equation}
H(xl_1,l_2,p)=\hat{H}(xl_1,l_2,p)\, 
              (2\pi) \delta ((l_2+xl_1-p)^2-m^2)\ ,
\label{e15}
\end{equation}
where $\hat{H}(xl_1,l_2,p)$ is given by the diagrams shown in 
Fig.~\ref{fig3}.  

In addition to the diagram in Fig.~\ref{fig1}, in general, 
we also need to consider the radiation diagrams shown in
Fig.~\ref{fig4}.  Similarly, the contribution of 
Fig.~\ref{fig4}a and Fig.~\ref{fig4}b 
can also be written in the same factorized form:
\begin{equation}
E \frac{d\sigma_{qq\rightarrow g}^{\rm rad}}{d^3 p} 
    \approx \frac{1}{2(2\pi)^3}\, \frac{1}{2s} \,
            \int \frac{dx}{x}\, 
             P_{l_1 \rightarrow k_1} (x, k_{1T}<p_T) \, 
             H_{i}(xl_1,l_2,p)
         +  O(\frac{\Lambda_{cutoff}^2}{p_T^2})  \, ,
\label{ae1}
\end{equation}
with $i=a, b$. Here $P_{l_1 \rightarrow k_1} (x, k_{1T}<p_T)$ is 
defined by Eq.~(\ref{e6}). $H_{a}(xl_1,l_2,p)$ and 
$H_{b}(xl_1,l_2,p)$ are the hard scattering parts from the diagrams in 
Fig.~\ref{fig4}a and Fig.~\ref{fig4}b, and they are represented by
Fig.~\ref{fig5}a and Fig.~\ref{fig5}b, respectively. 
With the contribution from Fig.~\ref{fig4}a and Fig.~\ref{fig4}b,
Eq.~(\ref{e5}) changes to
\begin{eqnarray}
E \frac{d\sigma_{qq\rightarrow g}}{d^3 p} 
   & \approx & 2\left(\frac{1}{2(2\pi)^3}\, \frac{1}{2s} \right)
            \int \frac{dx}{x}\, 
            P_{l_1 \rightarrow k_1} (x, k_{1T}<p_T) 
\nonumber \\
&\ & \quad\quad\quad \times 
          \left[ H(xl_1,l_2,p)
                +H_{a}(xl_1,l_2,p)
                +H_{b}(xl_1,l_2,p) 
                +\mbox{interference terms} \right]
\nonumber \\
&\ &  \quad\quad\quad 
         +  O(\frac{\Lambda_{cutoff}^2}{p_T^2})  \, ,
\label{ae3}
\end{eqnarray}
with the approximation $k_1 =xl_1 +O(k_{1T})$.

Feynman diagrams  shown in Fig.~\ref{fig3} and Fig.~\ref{fig5} form 
a gauge invariant subset for calculating the hard scattering parts,
$H(xl_1,l_2,p)$'s in Eq.~(\ref{ae3}).  
Fig.~\ref{fig5}a and  Fig.~\ref{fig5}b
are effectively the $s$-channel and $u$-channel diagrams for
the $gq\rightarrow gq$ partonic process.  Since we are only 
interested in the soft gluon limit, when $|t|<<s$, the contribution 
from these two diagrams can be neglected, in comparison to the 
contribution from the diagram in Fig.~\ref{fig3}.  In addition, under 
the collinear expansion $k_1\approx xl_1$, the gluon line which 
connects the partonic parts $P_{l_1 \rightarrow k_1}$ and 
$H(xl_1,l_2,p)$ is effectively on the mass-shell, and therefore, the 
partonic parts, $P$ and $H$ in Eq.~(\ref{e5}) are separately gauge 
invariant. 

Similar arguments can be held for the situation when $k_2^2 \sim 0$ or
$k_{2T}<< p_T$.  For example, after the collinear expansion for $k_2$, 
the contributions from diagrams shown in  Fig.~\ref{fig4}c and  
Fig.~\ref{fig4}d can be neglected in the soft gluon limit.  

Therefore, with the approximation of $k_1^2\sim 0$ (or $k_2^2\sim 0$) 
and the soft gluon limit, and a proper choice of the gauge, 
the dominant contribution for the 
partonic process $qq\rightarrow qqg$ comes from the diagram shown in
Fig.~\ref{fig1}. In the next section, 
we derive the leading contribution of the partonic process 
$qq\rightarrow qqg$ with the above approximation.

%%%%%%%%%%%%%% End of Section II %%%%%%%%%%%%%%%%%%%%%%%%%%%%%%%%%%%%%%%
%%%%%%%%%%%%%% Beginx Section III %%%%%%%%%%%%%%%%%%%%%%%%%%%%%%%%%%%%%%%
\section{Derivations}
\label{sec:3}

\begin{figure}
\epsfig{figure=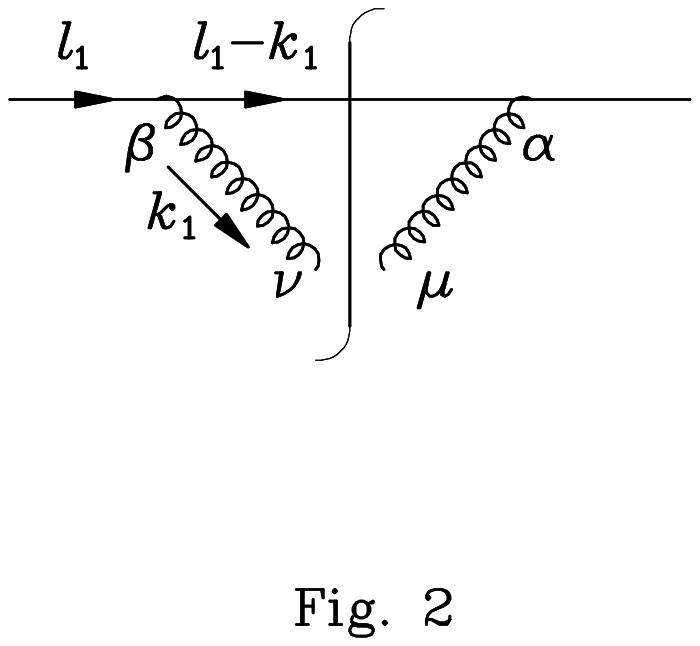,width=2.0in}
\caption{Feynman diagram for the splitting of $q\rightarrow g$.}
\label{fig2}
\end{figure}

\begin{figure}
\epsfig{figure=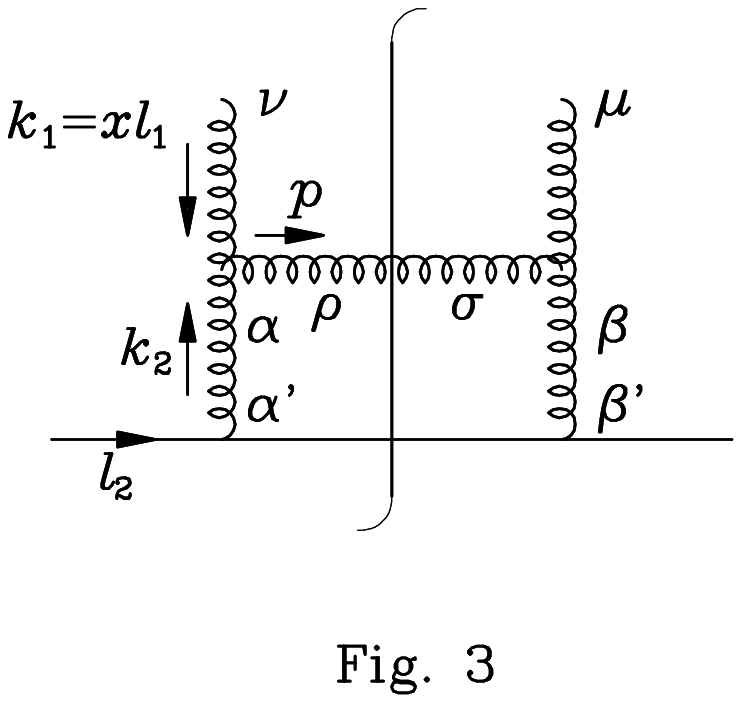,width=2.0in}
\caption{Leading Feynman diagram contributing to the hard scattering 
part $H(xl_1,l_2,p)$. }
\label{fig3}
\end{figure}

\begin{figure}
\epsfig{figure=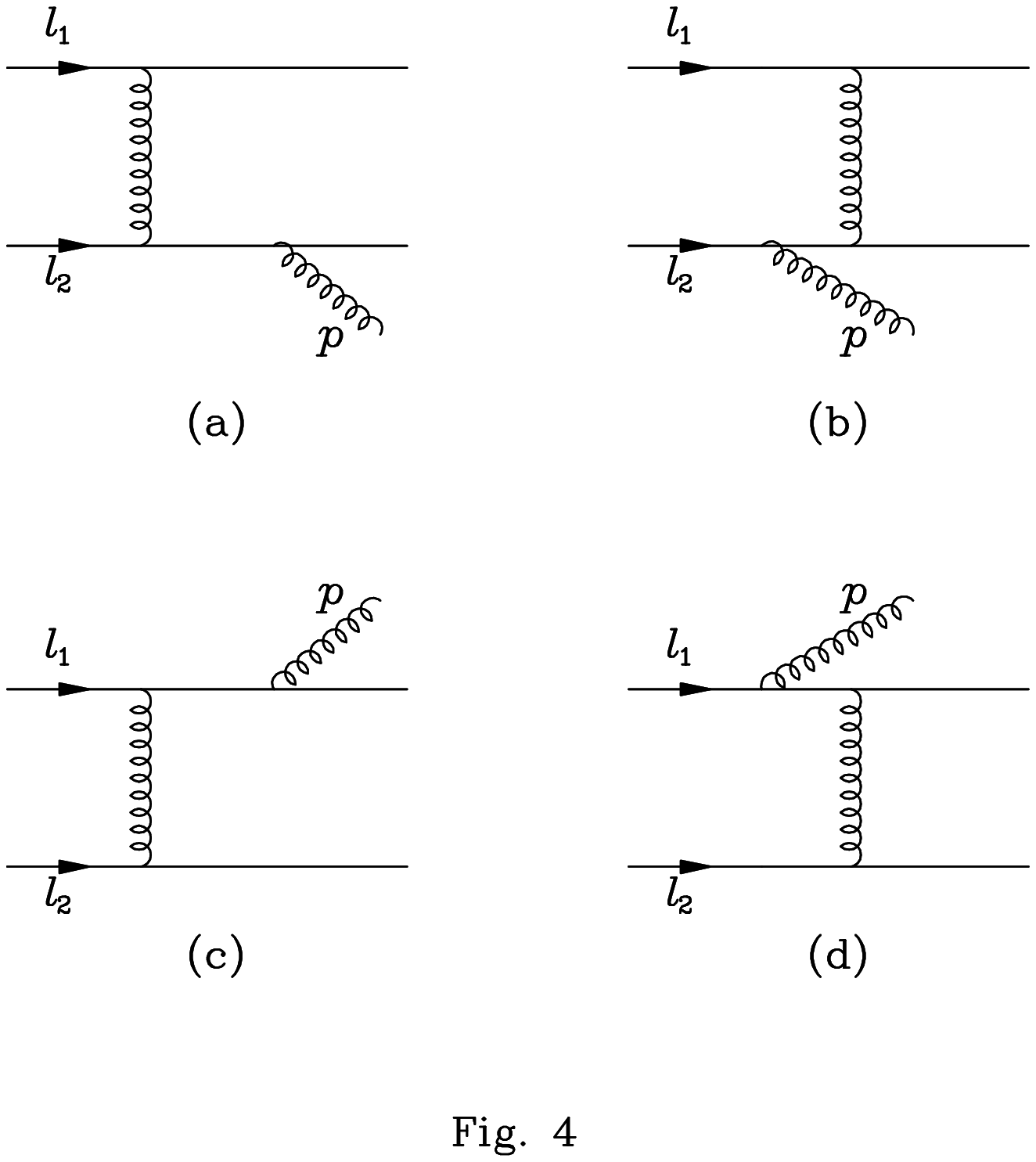,width=2.0in}
\caption{The rest Feynman diagrams to the process: 
$qq\rightarrow qqg$, in addition to the diagram in Fig.~1.}
\label{fig4}
\end{figure}

\begin{figure}
\epsfig{figure=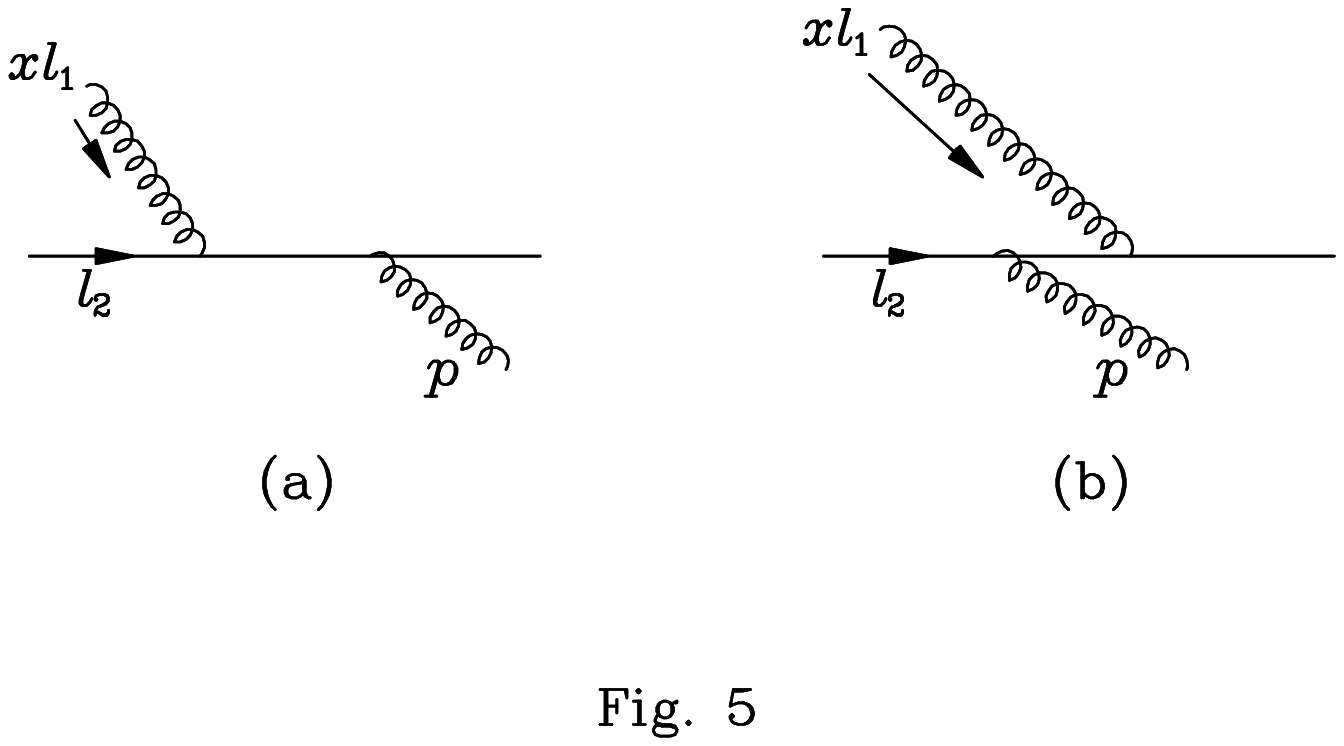,width=2.0in}
\caption{Feynman diagrams contributing to $H_a(xl_1,l_2,p)$ (a),
and $H_b(xl_1,l_2,p)$ (b). }
\label{fig5}
\end{figure}

Following the discussion in last section, we now turn to explicit 
calculation of the leading contribution to the gluon production 
from the partonic process 
$qq\rightarrow qqg$, shown in Fig.~\ref{fig1}. As shown in 
Eq.~(\ref{e5}), the leading contribution of this partonic process can be
factorized into two parts: $P_{l_1 \rightarrow k_1} 
(x, k_{1T}<p_T)$ and $H(xl_1,l_2,p)$. 
$P_{l_1 \rightarrow k_1} (x, k_{1T}<p_T)$ represents the splitting of 
the quark to the soft gluon with momentum fraction $x$, and 
$H(xl_1,l_2,p)$ represents the scattering between the gluon of momentum 
$xl_1$ and the other incoming quark of momentum $l_2$.  
$P_{l_1 \rightarrow k_1}(x, k_{1T}<p_T)$ and 
$H(xl_1,l_2,p)$ are represented by the diagrams in 
Fig.~\ref{fig2} and Fig.~\ref{fig3}, respectively. In the following 
derivation, we choose the center of mass frame, with 
\begin{equation} 
l_1=(l_{1+}, l_{1-},l_T)=(l_+, 0, 0)\ \ 
\mbox{and} \ l_2=(0,l_-,0)\, .
\label{e7}
\end{equation}
The definitions of the plus and minus components of the four momentum 
$p=(p_0, p_1, p_2, p_3)$ are:
\begin{equation}
p_+=\frac{p_0+p_3}{\sqrt 2}, \ \ p_-=\frac{p_0-p_3}{\sqrt 2}\, .
\label{e8}
\end{equation}
We also introduce two useful vectors, $n$ and $\bar{n}$ as:
\begin{equation}
n=(0,1,0_T), \ \ \ \bar{n}=(1,0,0_T) \ .
\label{e9}
\end{equation}

As we discussed above, $P_{l_1 \rightarrow k_1} 
(x, k_{1T}<p_T)$ and $H(xl_1,l_2,p)$ are separately gauge invariant. 
To derive the complete leading contribution, we choose 
$n \cdot A=0$ gauge to calculate 
$P_{l_1 \rightarrow k_1}(x, k_{1T}<p_T)$ and $\bar{n}\cdot A=0$ gauge 
for calculating the $H(xl_1,l_2,p)$. From Eq.~(\ref{e6}) and the 
diagram shown in Fig.~\ref{fig2}, we have in $n\cdot A=0$ gauge,
\begin{eqnarray}
 P_{l_1 \rightarrow k_1} (x, k_{1T}<p_T)  
&=& C_{q\rightarrow g}\,
    g^2 \int \frac{d^4 k_1}{(2\pi)^4} \, 
       x\,  \delta (x-\frac{k_1}{l_1})
    (2\pi) \delta ((l_1-k_1)^2-m^2) \nonumber \\
&\ & \times \frac{1}{2} {\rm Tr}
        (\gamma \cdot l_1 \gamma^{\alpha} \gamma \cdot (l_1-k_1)
         \gamma^{\beta}) 
        \frac{P_{\alpha \mu}(k_1)}{k_1^2} 
                \frac{P_{\beta \nu}(k_1)}{k_1^2}
                 (-g^{\mu \nu}),
\label{e10}
\end{eqnarray}
where $C_{q\rightarrow g}$ is the color factor.  In Eq.~(\ref{e10}),
the gluon polarization tensor is defined as
\begin{equation}
P_{\alpha \mu}(k_1)=-g_{\alpha \mu}  
      + \frac{k_{1 \alpha} n_\mu +n_\alpha k_{1 \mu}}{k_1 \cdot n}.   
\label{e11}
\end{equation}
The four dimension integral $d^4 k_1 =dk_{1+} dk_{1-} \cdot 
\pi\, dk_{1T}^2$. 
We can use the $\delta$-function  $\delta (x-\frac{k_1}{l_1})$ to fix 
$k_{1+}$, and use $\delta ((l_1-k_1)^2-m^2)$ to fix $k_{1-}$.
We have 
\begin{equation}
k_{1+}=x l_{+}, \ \  k_{1-}= -\frac{k_{1T}^2}{2l_{+} (1-x)}.
\label{e12}
\end{equation}
Substituting Eq.~(\ref{e12}) into Eq.~(\ref{e10}), and working out 
the trace, we obtain
\begin{eqnarray}
P_{l_1 \rightarrow k_1}(x,k_{1T}<p_T) 
 &=& C_{q\rightarrow g}\, 
        \frac{g^2}{8\pi^2} 
                    \frac{1+(1-x)^2}{x}
                    \int ^{p^2_T} _{\Lambda ^2 _{cutoff}}
                            d k_{1T}^2\, \frac{1}{k_{1T}^2}
\nonumber \\
 &=&\frac{N_c^2-1}{2N_c} \,  
    \left( \frac{g^2}{8\pi^2} \frac{1+(1-x)^2}{x} \right)
      \ell n \left(\frac{p_T^2}{\Lambda^2_{cutoff}}\right)\, .
\label{e13}
\end{eqnarray}

Choosing $\bar{n}\cdot A=0$ gauge, we derive the partonic scattering 
part of the $H(xl_1,l_2,p)$ defined in Eq.~(\ref{e15}) from 
the diagram shown in Fig.~\ref{fig3}, 
\begin{eqnarray}
\hat{H}(xl_1,l_2,p) 
&=& g^4\, \frac{1}{4}\,  d_{\mu \nu }\, {\rm Tr}
             (\gamma \cdot l_2 \gamma^ {\beta '}
              \gamma \cdot (l_2-k_2) \gamma ^{\alpha '})
\nonumber \\
& \ & \times
           P_{\rho \sigma }(p,\bar{n})
           \frac{P_{\beta \beta '}(k_2,\bar{n})}{k_2^2}
           \frac{P_{\alpha \alpha '}(k_2,\bar{n})}{k_2^2}
\nonumber \\
& \ & \times
\left[ (-2xl_1+p)^{\sigma } g^{\mu \beta } \right.
      + (-2p+xl_1)^{\mu } g^{\beta \sigma }        
    \left. + (p+xl_1)^{\beta } g^{\sigma \mu } \right]
\nonumber \\
&\ &  \times
\left[ (p-2xl_1)^{\rho } g^{\alpha \nu } \right.
      + (xl_1 +p)^{\alpha } g^{\nu \rho }
    \left. + (-2p+xl_1)^{\nu } g^{\rho \alpha } \right]\, ,
\label{e16}
\end{eqnarray}
where $d_{\mu \nu}$ is defined as
\begin{equation}
d_{\mu \nu}= -g_{\mu \nu} + n_{\mu} \bar{n}_{\nu} 
            + \bar{n}_{\mu} n_{\nu}\, .
\label{dmunu}
\end{equation}
In Eq.~(\ref{e16}), the gluon polarization tensors are given by
\begin{eqnarray}
P_{\rho \sigma }(p,\bar{n}) &=& -g_{\rho \sigma}  
      + \frac{p_{ \rho} \bar{n}_\sigma 
              +\bar{n}_\rho p_{\sigma}}{p \cdot \bar{n}}\, ,
\nonumber \\
P_{\beta \beta '}(k_2,\bar{n}) &=& -g_{\beta \beta '}  
      + \frac{k_{2 \beta} \bar{n}_{\beta '} 
           +\bar{n}_\beta k_{2 \beta '}}{k_2 \cdot \bar{n}}\, .
\label{epp}
\end{eqnarray}
Using the relations $k_2^2=(p-xl_1)^2$ and $p^2=2p_+ p_- -p_T^2=0$, 
we have
\begin{equation}
\frac{1}{k_2^2}=\frac{p_+}{xl_+}\, \frac{1}{p_T^2}
\label{e19}
\end{equation}
Substituting Eq.~(\ref{e19}) into Eq.~(\ref{e16}), and after some algebra, 
we obtain 
\begin{equation}
\hat{H}(xl_1,l_2,p)
=4g^4 \left( \frac{p_+}{xl_+} \right) ^2 \frac{1}{p_T^4}
        \left[ (xs-2xl_+p_-)^2+(2xl_+p_-)(2p_+p_-) \right]\, ,
\label{e20}
\end{equation}
where $s=(l_1+l_2)^2=2l_{+}l_{-}$ is the total invariant 
mass squared of the two incoming quarks.  
Substituting Eq.~(\ref{e20}) into Eq.~(\ref{e16}), and after taking into 
account of the color factor $1/2$, we obtain the hard scattering 
function $H(xl_1,l_2,p)$ as
\begin{eqnarray}
H(xl_1,l_2,p) &=& (2\pi) 4g^4 \left(\frac{1}{2}\right)
        \left( \frac{p_+}{xl_+} \right) ^2 \frac{1}{p_T^4}
       \frac{1}{s-2l_+p_-} 
       \delta (x-\frac{2p_+l_-}{s-2l_+p_-} ) \nonumber \\
&\ & \times
       \left[ (xs-2xl_+p_-)^2+(2xl_+p_-)(2p_+p_-) \right]\, .
\label{e21}
\end{eqnarray}

Combining Eq.~(\ref{e13}), Eq.~(\ref{e21}) and Eq.~(\ref{e5}), we obtain
\begin{eqnarray}
E \frac{d\sigma_{qq\rightarrow g}}{d^3p} 
&=& \frac{g^6}{(2\pi)^4} 
               \left(\frac{1}{2N_c}\right)^2 N_c (N_c^2-1)
\nonumber \\
&\ & \times
     \int dx \delta (x-\frac{2p_+l_-}{s-2l_+p_-} ) (1+(1-x)^2)
      \left(\frac{p_+}{xl_+}\right)^2 
      \frac{1}{s(s-2l_+p_-)}    \nonumber \\
&\ & \quad \times
  \left[ (s-2l_+p_-)^2+\frac{2l_+p_-}{x} (2p_+p_-) \right]
   \left( \frac{1}{p_T^4} \right) 
    \ell n \left(\frac{p_T^2}{\Lambda ^2 _{cutoff}}\right)\, .
\label{e22}
\end{eqnarray}

We define the soft gluon limit as
\begin{equation}
\frac{p_{-}}{l_{-}} \ll 1 \,  
\quad \mbox{and} \quad
\frac{p_{+}}{l_{+}} \ll 1 \, .
\label{e23}
\end{equation}
At this soft gluon limit, we have $s=2l_{+}l_{-}>>2l_{+}p_{-}$; and 
from the $\delta$-function in Eq.~(\ref{e22}), we have 
\begin{equation}
x \approx \frac{p_+}{l_+} \ll 1, \quad \mbox{and} \quad  
1+(1-x)^2 \approx 2 \, .
\label{e24}
\end{equation}
Substituting Eq.~(\ref{e24}) into Eq.~\ref{e22}, 
and taking the soft gluon limit, we obtain
\begin{equation}
E \frac{d\sigma_{qq\rightarrow g}}{d^3p} 
  = \frac{2g^6}{(2\pi)^4}
    \left(\frac{1}{2N_c}\right)^2 
    N_c (N_c^2-1)\left( \frac{1}{p_T^4} \right) 
    \ell n \left(\frac{p_T^2}{\Lambda ^2 _{cutoff}}\right)\, .
\label{e25}
\end{equation}
Defining $y=\frac{1}{2} \ell n(\frac{E+p_z}{E-p_z})$, 
we can rewrite  the cross section in terms of variable $y$ 
\begin{equation}
\frac{d\sigma_{qq\rightarrow g}}{dy d^2 p_T}
  = \frac{2g^6}{(2\pi)^4}
    \left(\frac{1}{2N_c}\right)^2 
    N_c (N_c^2-1)\left( \frac{1}{p_T^4} \right) 
    \ell n \left(\frac{p_T^2}{\Lambda ^2 _{cutoff}}\right)\, .
\label{e26}
\end{equation}
This is our final result. Eq.~(\ref{e26}) shows the same $p_T^2$ 
dependence as the result obtained by Kovner, McLerran, and 
Weigert \cite{KMH}. The difference between 
Eq.~(\ref{ae0}) is the factor of $N_q^2$.
Eq.~(\ref{e26}) is obtained by calculating the
leading contribution of the subprocess $qq\rightarrow qqg$, 
for which $N_q$ effectively equals to one. If we consider
the total number of the quarks in the charge sources of both sides, 
we need to multiply $N_q^2 $ to Eq.~(\ref{e26}), and our result 
reproduces the result obtained by KMW.

%%%%%%%%%%%%%% End of Section III %%%%%%%%%%%%%%%%%%%%%%%%%%%%%%%%%%%%%%
%%%%%%%%%%%%%% Begin Section IV %%%%%%%%%%%%%%%%%%%%%%%%%%%%%%%%%%%%%%%%
\section{Summary and Discussions}
\label{sec:4}

In this section we discuss the similarities and differences between 
KMW's result, Eq.~(\ref{e1}) or Eq.~(\ref{ae0}), which was obtained 
in McLerran-Venugopalan formalism, and our result, Eq.~(\ref{e26}), 
which was obtained in the conventional perturbative QCD formalism 
at the leading logarithmic approximation.  

Our result can be reexpressed in terms of the usual factorized 
cross section in perturbative QCD. 
When we consider the collision between two nuclei, we can treat the two 
incoming quarks in Fig.~\ref{fig1} as coming from two nuclei respectively. 
In this picture, the number of the valence quark $N_q$ is replaced by the
quark distribution in the nuclei.
The cross section in Eq.~(\ref{e5}) (or equivalently Eq.~(\ref{e25})) 
is just the partonic cross section for the collision between two quarks. 
In terms of the parton model, 
the cross section between the two nuclei $A$ and $B$ can be 
expressed in the following form:
\begin{equation}
E \frac{d\sigma_{AB\rightarrow g}}{d^3p} = 
         \int dz_1\, dz_2\, f_{q/A}(z_1)\, f_{q/B}(z_2)\,
         E \frac{d\sigma_{qq\rightarrow g}}{d^3 p}\, .
\label{ae4}
\end{equation}
Here $z_1$ and $z_2$ are the momentum fractions of the quarks, and 
$f_{q/A}(z_1)$ and  $f_{q/B}(z_2)$ are the quark distributions (or quark 
number densities) of the two nuclei.  If we denote
$p_A$ and $p_B$ as the momenta for the two nuclei respectively, then
$z_1=l_1/p_A$  and $z_2=l_2/p_B$. Substituting Eq.~(\ref{e5}) into
Eq.~(\ref{ae4}), we have
\begin{eqnarray}
E \frac{d\sigma_{AB\rightarrow g}}{d^3p} & \approx & 
    \frac{1}{2(2\pi)^3}\, \frac{1}{2S} \, 
    \int \frac{dz_1}{z_1} \, \frac{dz_2}{z_2} \left[
    \int \frac{dx_1}{x_1}\, f_{q/A}(z_1) f_{q/B}(z_2)
    P_{l_1 \rightarrow k_1} (x_1, k_{1T}<p_T) \, H(x_1l_1,l_2,p) \right.
\nonumber \\
&\ & \quad \quad \quad \quad + \left.
    \int \frac{dx_2}{x_2}\, f_{q/A}(z_1) f_{q/B}(z_2)
    P_{l_2 \rightarrow k_2} (x_2, k_{2T}<p_T) \, H(l_1,x_2l_2,p) \right]
\label{ae5}
\end{eqnarray}
where the overall factor 2 in Eq.~(\ref{e5}) is now represented by 
the two terms, and $S=(p_A+p_B)^2\approx 2p_A \cdot p_B$.  
In Eq.~(\ref{ae5}), $x_i=k_i/l_i$ with $i=1,2$, 
and $l_1=z_1p_A$ and $l_2=z_2p_B$. If we denote the 
momentum fraction of gluon $k_1$ with respect to $p_A$ 
as $z_1'=k_1/p_A$, and $k_2$ with respect to $p_B$ 
as $z_2'=k_2/p_B$, we can rewrite Eq.~(\ref{ae5}) in terms of 
$z_1'$ and $z_2'$:
\begin{eqnarray}
E \frac{d\sigma_{AB\rightarrow g}}{d^3p} & \approx &
    \frac{1}{2(2\pi)^3}\, \frac{1}{2S} \,
    \int \frac{dz_1'}{z_1'} \, \frac{dz_2'}{z_2'}  
\nonumber \\
&\ & {\hskip 0.1in}
\times \left\{  
     \left[ \int \frac{dz_1}{z_1}\, f_{q/A}(z_1)\,
            P_{l_1 \rightarrow k_1} (z_1'/z_1, k_{1T}<p_T) \right]
    \right.
\nonumber \\
&\ & {\hskip 0.5in} \times 
    \left( \int \frac{dz_2}{z_2}\, f_{q/B}(z_2)\,
           \delta\left(1-\frac{z_2'}{z_2}\right) \right) 
\nonumber \\
&\ & {\hskip 0.3in} +\,
    \left( \int \frac{dz_1}{z_1}\, f_{q/A}(z_1)\,
           \delta\left(1-\frac{z_1'}{z_1}\right) \right)
\nonumber \\
&\ & {\hskip 0.5in} \times \left.
    \left[ \int \frac{dz_2}{z_2}\, f_{q/B}(z_2)\,
           P_{l_2 \rightarrow k_2} (z_2'/z_2, k_{2T}<p_T) \right]
    \right\}
        H(z_1'p_A,z_2'p_B,p)
\label{ae6} \\
&\approx &
    2\left(\frac{1}{2(2\pi)^3}\, \frac{1}{2S} \right)
    \int \frac{dz'}{z'} \, \frac{dz}{z}\,   
     \left[ \int \frac{dz_1}{z_1}\, f_{q/A}(z_1)\,
            P_{l_1 \rightarrow k_1} (z'/z_1, k_{1T}<p_T) \right]
\nonumber \\
&\ & {\hskip 1.5in} \times
     f_{q/B}(z)\, H(z'p_A,zp_B,p)\, .
\label{ae6p}
\end{eqnarray}
In obtaining Eq.~(\ref{ae6p}), we used the fact that the 
partonic scattering part $H(k_1,k_2,p)$ in 
Eq.~(\ref{ae6}) is symmetric under the exchange of the 
$k_1$ and $k_2$ at the soft gluon limit.

According to the QCD factorization theorem \cite{Factorization}, 
we see that the part inside the square
brackets is actually the gluon distribution from nuclei $A$ 
(or $B$) at 
the factorization scale $\mu_F^2=p_T^2$, with only the quark 
splitting function \cite{Field}, 
\begin{eqnarray}
f_{g/A}(z_1',\mu_F^2=p_T^2) &=& 
        \int \frac{dz_1}{z_1}\, f_{q/A}(z_1) 
        P_{l_1 \rightarrow k_1} (z_1'/z_1, k_{1T}<p_T) 
\nonumber \\
&\ & +   \mbox{term from gluon splitting}.
\label{ae7}
\end{eqnarray}
Using Eq.~(\ref{ae7}), we can then reexpress Eq.~(\ref{ae6}) as:
\begin{eqnarray}
E \frac{d\sigma_{AB\rightarrow g}}{d^3p}  & \approx &
        \frac{1}{2(2\pi)^3}\, \frac{1}{2S} \,
        \int \frac{dz'}{z'} \, \frac{dz}{z}  
         \left[ f_{g/A}(z',\mu_F^2=p_T^2) f_{q/B}(z) 
                H(z'p_A,zp_B,p) \right.
\nonumber \\
&\ & {\hskip 1.3in} \left. 
              + f_{q/A}(z) f_{g/B}(z',\mu_F^2=p_T^2) 
                H(zp_A,z'p_B,p) \right]
\, ,
\label{ae8}
\end{eqnarray} 
which is the factorized formula for  two-to-two subprocesses in 
the conventional perturbative QCD for the nucleus-nucleus collisions.
In KMW formalism, only the valence quark color charge was used 
as the source of the classical charge of colors.  As a result,  
the gluon splitting term in Eq.~(\ref{ae7}) is neglected for the 
distribution $f_{g/A}$.

Our discussions above show that the soft gluon distribution in heavy 
ion collisions obtained in KMW's approach can be 
understood by calculating the partonic process 
$qq \rightarrow qqg$.  To relate KMW's result to the factorized 
formula in the conventional perturbative QCD, we need to:  
(1) replace the charge density for the classical color charge 
$\mu^2$ (or equivalently $N_q$) by the valence quark distributions
of the nuclei; (2) absorb the logarithm 
$\ell n (p_T^2/\Lambda^2_{cutoff})$ into 
one of the valence quark distributions, which effectively becomes the 
gluon distribution of one of the initial nuclei.
From Eq.~(\ref{ae6}), we identify that the  
$\ell n (p_T^2 /\Lambda ^2 _{cutoff} )$ factor in  
Eq.~(\ref{e1}) (or in Eq.~(\ref{ae0})) comes from the logarithm 
of the splitting of the incoming quark to the soft gluon in 
Fig.~\ref{fig2}. In terms of the conventional QCD factorization theorem 
\cite{Factorization}, such logarithm is normally factorized into the
distributions of the collinear gluons inside the incoming hadrons, as 
demonstrated in Eq.~(\ref{ae7}), and
the logarithmic dependence of the distributions is a direct result of
solving the DGLAP evolution equations \cite{GLAP}.  

From the above comparison, we conclude that 
by solving the classical Yang-Mills Equation to the second order in 
iteration, KMW's result reproduces the result of conventional 
perturbative QCD at the leading logarithmic approximation, with 
the convolution over the parton number densities inside the nuclei 
replaced by the effective numbers of the valence quarks.
The logarithmic dependence shown in KMW's result basically 
describes the logarithmic DGLAP evolution of the quark distributions. 
However, in addition to the valence quarks, the glue field 
at small $x$ can be produced by {\it all} flavor partons that have 
larger momenta. 

The McLerran-Venugopalan formalism was later further developed by 
Ayala, Jalilian-Marian, Kovner, McLerran, Leonidov, Venugopalan, 
and Weigert \cite{JKMW,JKLW}. 
The major improvement to the McLerran-Venugopalan 
model is to include the harder gluons into the charge density $\mu^2$
and treat the charge source as an extended distribution which depends 
on the rapidity \cite{JKMW}. These improvements lead to 
the ``renormalization'' of the charge density. It was showed that 
the renormalization group 
equation for the charge density can be reduced to the BFKL 
equations \cite{BFAP} in some appropriate limits \cite{JKLW}.   
It will be very interesting to see if KMW's approach, after including
higher orders of iteration, can 
show the parton recombination \cite{recom} 
and other non-perturbative effects which are not 
apparent in the normal perturbative calculation.

%%%%%%%%%%%%%% End of Section IV %%%%%%%%%%%%%%%%%%%%%%%%%%%%%%%%%%%%%%%
%%%%%%%%%%%%%% Begin Acknoledgement %%%%%%%%%%%%%%%%%%%%%%%%%%%%%%%%%%%%
\section*{Acknowledgment}

I thank Miklos Gyulassy for stimulating discussions in the 
early stage of this work. I also thank Al Mueller for very helpful 
comments when I presented this work in an informal seminar at Columbia 
University. I thank Jianwei Qiu for very helpful discussions. I am 
grateful to Ed Berger, Susan Gardner, George Sterman, and Bin Zhang 
for their encouragement and support. I especially thank George Sterman 
for his reading of the manuscript and helpful suggestions. This work 
is supported in part by the U.S. Department of Energy under Grant Nos. 
DE-FG02-93ER40764 and DE-FG02-96ER40989.
%%%%%%%%%%%%%% End of Acknoledgement %%%%%%%%%%%%%%%%%%%%%%%%%%%%%%%%%%%
%%%%%%%%%%%%%% Begin References %%%%%%%%%%%%%%%%%%%%%%%%%%%%%%%%%%%%%%%%

%%%%%%%%%%%%%% End of References %%%%%%%%%%%%%%%%%%%%%%%%%%%%%%%%%%%
%%%%%%%%%%%%%%%%% Begin Figure Captions %%%%%%%%%%%%%%%%%%%%%%%%%%%%%%%

%%%%%%%%%%%%%%%%%%% End of Figure Captions %%%%%%%%%%%%%%%%%%%%%%%%%%
\end{document}